\documentclass[aps,twocolumn]{revtex4}
\usepackage{epsfig}
\begin{document}
\def\I.#1{\it #1}
\def\B.#1{{\bbox#1}}
\def\C.#1{{\cal  #1}}
\title{Multifractal Structure of the Harmonic Measure of
Diffusion Limited Aggregates}
\author {Mogens H. Jensen$^*$, Anders Levermann$^{**}$, Joachim
Mathiesen$^{*}$, and Itamar Procaccia$^{**}$}
\address{$^{*}$ The Niels Bohr Institute, 17 Blegdamsvej, Copenhagen,
Denmark\\
$^{**}$Department of~~Chemical Physics, The Weizmann Institute of
Science, Rehovot 76100, Israel}

%
%

\begin{abstract}
The method of iterated conformal maps allows to study the 
harmonic measure of Diffusion Limited Aggregates with unprecedented
accuracy. We employ this method to explore the multifractal properties
of the measure, including the scaling of the measure in the deepest fjords
that were hitherto screened away from any numerical probing. We resolve probabilities 
as small as $10^{-35}$, and present an accurate determination of the
generalized dimensions and the spectrum of singularities. We show that the generalized
dimensions $D_q$ are infinite for $q<q^*$, where $q^*$ is of the order of $-0.2$. In the
language of $f(\alpha)$ this means that
$\alpha_{\rm max}$ is finite. The $f(\alpha)$ curve loses analyticity (the phenomenon
of ``phase transition") at $\alpha_{\rm max}$ and a finite value of $f(\alpha_{\rm max})$.
We consider the geometric structure of the regions that support the lowest parts of 
the harmonic measure, and thus offer an explanation for the phase transition,
rationalizing the value of $q^*$ and $f(\alpha_{\rm max})$. We thus offer 
a satisfactory physical
picture of the scaling properties of this multifractal measure. 
\end{abstract}
\maketitle

\section{introduction}
Multifractal measures are normalized distributions lying upon fractal sets.
Such measures appear naturally in a variety of nonlinear-physics context, the
most well studied being natural measures of chaotic dynamcial systems
\cite{78Rue,83GP,85ER}. Other well studied examples are the voltage distribution of random
resistor networks \cite{85RTBT,85ARC}. In this paper we address the harmonic measure of
Diffusion Limited Aggregates \cite{81WS}, which is the probability measure for a random
walker coming from infinity to hit the boundary of the fractal cluster. This was one of
the earliest multifractal measures to be studied in the physics literature \cite{86HMP}, 
but the elucidation of its properties was made difficult by the extreme
variation of the probability to hit the tips of a DLA versus hitting the deep fjords. 
With usual numerical techniques it is quite impossible to estimate accurately the extremely
small probabilities to penetrate the fjords. Contrary to harmonic measures of
conformally invariant fractals like random walks and percolation clusters whose multifractal
properties can be solved exactly \cite{00Dup,01Has}, the present multifractal measure posed
stubborn barriers to mathematical progress.

The multifractal properties of fractal measures in general, and of
the harmonic measure of DLA in particular, are conveniently
studied in the context of the generalized dimensions $D_q$, and the
associated $f(\alpha)$ function \cite{83HP,86HJKSP}. The simplest definition of the
generalized dimensions
is in terms of a uniform covering of the boundary of a DLA cluster
with boxes of size $\ell$, and measuring the probability for a random
walker coming from infinity to hit a piece of boundary which belongs to
the $i$'th box.
Denoting this probability by $P_i(\ell)$, one considers \cite{83HP}
\begin{equation}
D_q \equiv \lim_{\ell \to 0}\frac{1}{q-1}\frac{\log\sum_i
P_i^q(\ell)}{\log\ell} \ ,
\end{equation}
where the index $i$ runs over all the boxes that contain a piece of the boundary.
The limit $D_0\equiv \lim_{q\to 0^+} D_q$ is the fractal, or box dimension of the cluster.
$D_1\equiv \lim_{q\to 1^+} D_q$ and
$D_2$ are the well known information and correlation 
dimensions respectively \cite{83GP,56BR,82Far}. It is
well established by now \cite{86HJKSP} that the existence of an interesting spectrum of
values $D_q$ is related to the probabilities $P_i(\ell)$ having a spectrum of
``singularities" in the
sense that $P_i(\ell) \sim \ell^\alpha$
with $\alpha$ taking on values from a range $\alpha_{\rm min}\le
\alpha\le \alpha_{\rm max}$.
The frequency of observation of a particular value of $\alpha$ is
determined by the the function
$f(\alpha)$ where (with $\tau(q)\equiv (q-1)D_q$)
\begin{equation}
f(\alpha) = \alpha q(\alpha)-\tau\Big(q\left(\alpha\right)\Big)\ ,\quad
\frac{\partial \tau(q)}{\partial q}=\alpha(q) \ .
\end{equation}

The understanding of the multifractal properties and the associated
$f(\alpha)$ spectrum of DLA clusters have been a long standing
issue. Of particular interest are the values of the minimal
and maximal values, $\alpha_{min}$ and $\alpha_{max}$, relating
to the largest and smallest growth probabilities, respectively.

The minimal value of $\alpha$ is relatively easy to estimate, since 
it has to do with the scaling of the harmonic measure near the most probable tip.
While the often cited Turkevich-Scher conjecture \cite{85TS} that $\alpha_{\rm min}$
satisfies the scaling relation $D_0=1+\alpha_{\rm min}$ is probably not exact,
it comes rather close to the mark. On the other hand, the
maximal value of $\alpha$ is a much more subtle issue. As a DLA cluster grows the
large branches screen the deep fjords more and more and the probability for a random walker
to get into these fjords (say around the seed of the cluster) becomes
smaller and smaller. A small growth probability corresponds  to a
large value of $\alpha$.
Previous literature hardly agrees about the actual value of
$\alpha_{max}$.
Ensemble averages of the harmonic measure
of DLA clusters indicated a rather large value of $\alpha_{max} \sim 8$
\cite{averages}.
In subsequent experiments on non-Newtonian fluids \cite{Nittmann}
and on viscous fingers \cite{viscous},
similar large values of $\alpha_{max}$ were also observed.
These numerical and experimental indications of a very large value
of $\alpha_{max}$ led to a conjecture that, in the limit
of a large, self-similar cluster some fjords will be exponentially
screened and thus causing $\alpha_{max} \to \infty$ \cite{Bohr}.

If indeed $\alpha_{max} \to \infty$, this can be interpreted
as a phase transition \cite{Predrag} (non-analyticity) in the $q$
dependence
of $D_q$, at a value of $q^*$ satisfying $q^*\ge 0$. If the transition takes
place for a value
$q^* < 0$ then $\alpha_{max}$ is finite. Lee and Stanley \cite{88LS}
proposed that $\alpha_{max}$
diverges like $R^2/\ln{R}$ with $R$ being the radius of the cluster.
Schwarzer et al. \cite{schwarzer} proposed that
$\alpha_{max}$ diverges only logarithmically in the number of added
particles.
Blumenfeld and Aharony \cite{amnon} proposed that channel-shaped fjords
are important and proposed that $\alpha_{max} \sim {{M^x} \over {ln~M}}$
where $M$ is the mass of the cluster;
Harris and Cohen \cite{90HC}, on the other hand, argued that straight
channels might be so
rare that they do not make a noticeable contribution, and
$\alpha_{max}$ is finite, in agreement with Ball and Blumenfeld who
proposed \cite{Ball}
that $\alpha_{max}$ is bounded. Obviously, the issue was not
quite settled. The difficulty
is that it is very hard to estimate the smallest growth probabilities
using models or direct numerical simulations.

In a recent Letter \cite{01DJLMP} we used the method of iterated conformal maps to offer an
accurate determination of the probability for the rarest events. 
The main result that was announced was
that $\alpha_{max}$ exists and the phase transition occurs at a $q$
value that is slightly negative. In the present
paper we discuss the results in greater detail, and offer additional
insights to the geometric interpretation of the phase transition.
In Sect. 2 we summarize briefly the method of iterated conformal maps
and explain how it is employed to compute the harmonic measure of
DLA with unprecedented accuracy. In Sect. 3 we perform the multifractal
analysis and present the calculation of the $f(\alpha)$ curve. In Sect. 4
we discuss a complementary point of view of the scaling properties of
the rarest regions of the measure, to achieve in Sect. 5 a
geometric interpretation of the phase transition. Sect. 6 offers a short discussion.
\section{Accurate calculation of the Harmonic Measure} 
\subsection{DLA via iterated conformal maps}
Consider a DLA of $n$ particles and denote the boundary of the cluster
by $z(s)$ where $s$ is an arc-length paramterization. Invoke now a
conformal map $\Phi^{(n)}(\omega)$ which maps the exterior of the unit 
circle in the mathematical plane $\omega$ onto the complement of the
cluser of $n$ particle in the $z$ plane. On the unit circle $e^{i\theta}$
the harmonic measure is uniform, $P(\theta)d\theta=d\theta/2\pi $. 
The harmonic measure of an element $ds$ on the cluster in the physical
space is then determined as
\begin{equation}
P(s)ds \sim \frac{ds}{|\Phi^{'(n)}(e^{i\theta})|} \ . \label{defharm}
\end{equation}
where $\Phi^{(n)}(e^{i\theta})=z(s)$. Note that in electrostatic parlance 
$1/|\Phi^{'(n)}(\omega)|$ is the electric field 
at the position $z=\Phi^{(n)}(\omega)$. Thus in principle, if we
can have an accurate value of the conformal map $\Phi^{(n)}(\omega)$
for all values $\omega=e^{i\theta}$ we can compute the harmonic
measure with desired precision. We will see that this is easier
said than done, but nevertheless this is the basic principle
of our approach.

We thus need to find the conformal map $\Phi^{(n)}(\omega)$. An excellent
method for this purpose was developed in a recent series of papers
\cite{98HL,99DHOPSS,00DFHP}.
The map $\Phi^{(n)}(w)$ is
made from compositions of elementary maps $\phi_{\lambda,\theta}$,
\begin{equation}
\Phi^{(n)}(w) = \Phi^{(n-1)}(\phi_{\lambda_{n},\theta_{n}}(w)) \ ,
\label{recurs}
\end{equation}
where the elementary map $\phi_{\lambda,\theta}$ transforms the unit
circle to a circle with a semi-circular ``bump" of linear size $\sqrt{\lambda}$ around
the point $w=e^{i\theta}$. We use below the same map $\phi_{\lambda,\theta}$ that
was employed in \cite{98HL,99DHOPSS,00DFHP,00DP,00DLP}.
With this map $\Phi^{(n)}(w)$ adds on a semi-circular
new bump to the image of the unit circle under $\Phi^{(n-1)}(w)$. The
bumps in the $z$-plane simulate the accreted particles in
the physical space formulation of the growth process. Since we want
to have {\em fixed size} bumps in the physical space, say
of fixed area $\lambda_0$, we choose in the $n$th step
\begin{equation}
\lambda_{n} = \frac{\lambda_0}{|{\Phi^{(n-1)}}' (e^{i \theta_n})|^2} \ .
\label{lambdan}
\end{equation}
The recursive dynamics can be represented as iterations
of the map $\phi_{\lambda_{n},\theta_{n}}(w)$,
\begin{equation}
\Phi^{(n)}(w) =
\phi_{\lambda_1,\theta_{1}}\circ\phi_{\lambda_2,\theta_{2}}\circ\dots\circ
\phi_{\lambda_n,\theta_{n}}(\omega)\ . \label{comp}
\end{equation}
It had been demonstrated before that this method represents DLA
accurately, providing many analytic insights that are not available
otherwise \cite{00DP,00DLP}. 

\subsection{Computing the harmonic measure}
In terms of computing the harmonic measure we note the
close relationship between Eqs.(\ref{defharm}) and (\ref{lambdan}).
Clearly, moments of the harmonic measure can be computed from
moments of $\lambda_n$. For our purposes here we quote a result
established in \cite{99DHOPSS}, which is
\begin{equation}
\langle \lambda^q_n \rangle\equiv
(1/2\pi)\int_0^{2\pi}\lambda^q_n(\theta) d\theta
 \sim n^{-2qD_{2q+1}/D} \ . \label{lamDq}
\end{equation}
To compute $\tau(q)$ we rewrite this average as
\begin{equation}
\label{mean}
\langle \lambda^q_n \rangle=
\int ds  \left|\frac{d\theta}{ds}\right| ~\lambda^q_n(s)=\int ds
\frac{\lambda^{q+1/2}(s)}{\sqrt{\lambda_0}} \ ,
\end{equation}
where $s$ is the arc-length of the physical boundary of the cluster. In
the last equality we used the fact that
$|d\theta/ds|=\sqrt{\lambda_n/\lambda_0}$. We stress at this point that
in order to measure
these moments for $q\le 0$ we {\em must} go into arc-length representation.

To make this crucial point clear
we discuss briefly what happens if one attempts to compute the moments
from the definition
(\ref{lamDq}). Having at hand the conformal map
$\Phi^{(n)}(e^{i\theta})$,  one
can choose randomly as many points on the unit circle $[0,2\pi]$ as one
wishes,
obtain as many  (accurate) values of $\lambda_n$, and try to compute the
integral
as a finite sum. The problem is of course that using such an approach
{\em the
fjords are not resolved}. To see this we show in Fig.1 panel a the
region
of a typical cluster of 50 000 particles that is being  visited by a
random search on the unit circle, with 50 000 samples. Like in direct simulations using
random walks, the rarest events are not probed, and no
serious conclusion regarding the phase transition is possible.
\begin{figure}
\epsfxsize=6truecm
\epsfbox{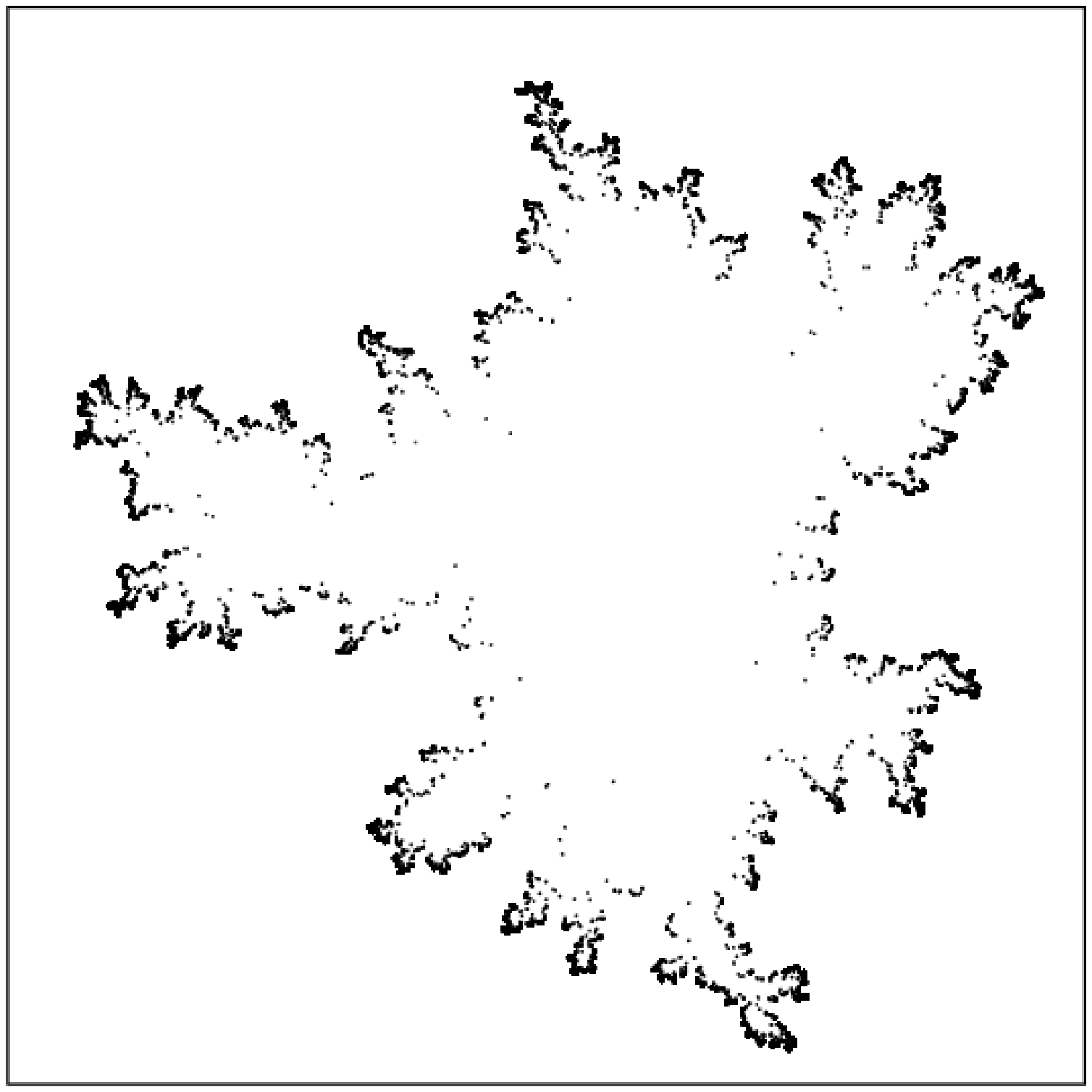}
\epsfxsize=6truecm
\epsfbox{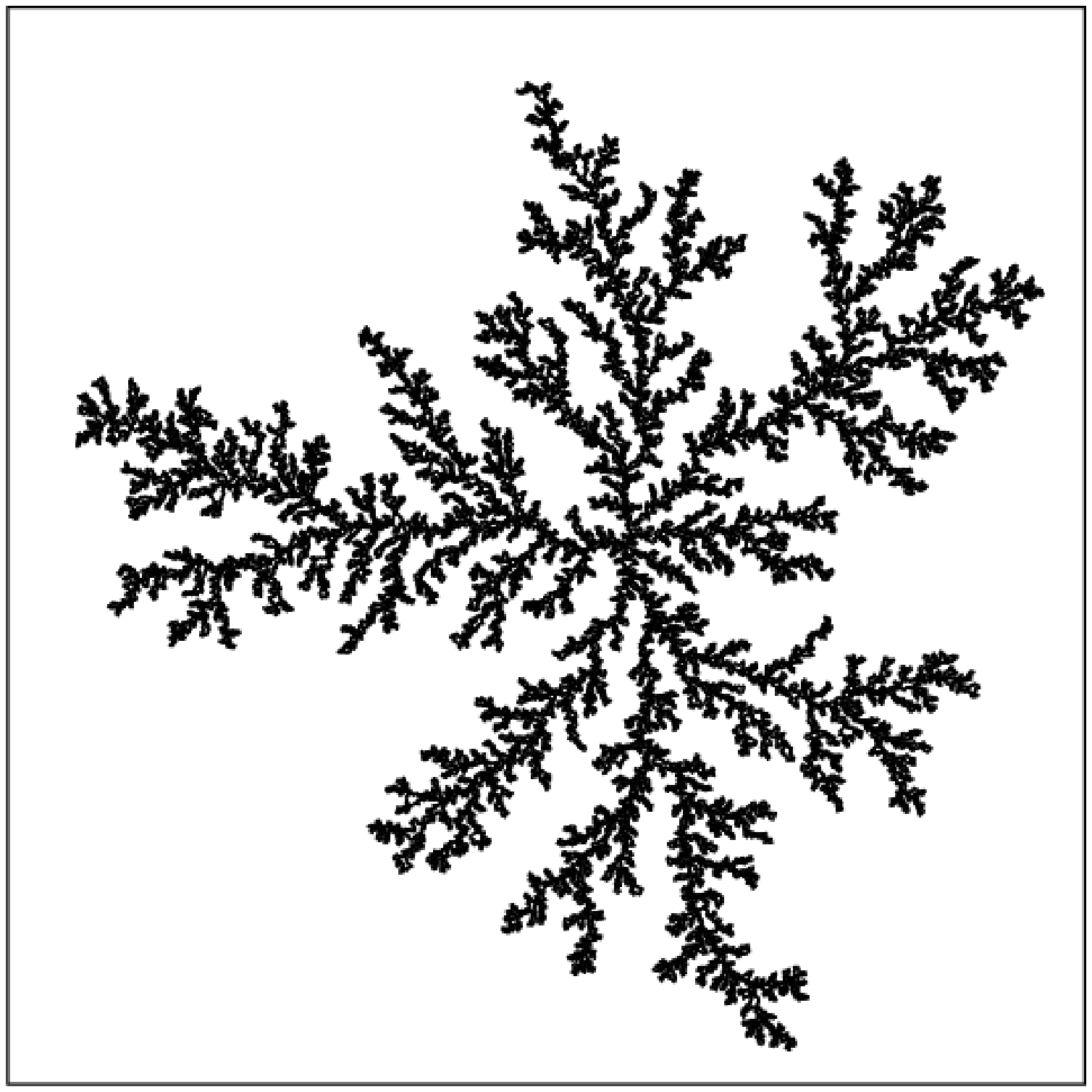}
\caption{panel a: the boundary of the cluster probed by a random search
with respect to the harmonic measure. Panel b: the boundary of the
cluster probed by the
present method.}
\label{patterns}
\end{figure}
Another method that cannot work is to try to compute $\langle \lambda_n^q \rangle$ by
sampling on the arc-length in a naive way. The reason is that the inverse
map $[\Phi^{(n)}]^{-1}(s)$ cannot resolve $\theta$ values that belong
to deep fjords. As the growth proceeds, reparametrization squeezes
the $\theta$ values that map to fjords
into minute intervals, below the computer numerical resolution.
To see this, consider the following estimate of the resolution we can
achieve in the physical space,
\begin{equation}
\Delta \theta=\frac{\Delta
  s}{\left|\Phi^{(n)'}\right|}=\sqrt\lambda_n\frac{\Delta s}{\sqrt \lambda_0} \ ,
\end{equation}
or equivalently
\begin{equation}
\frac{\Delta s}{\sqrt \lambda_0}=\frac{\Delta \theta}{\sqrt \lambda_n} \ .
\end{equation}
On the left hand side we have the resolution in the physical space relative to
the fixed linear size of the particles. With double
precission numerics we can resolve values of $\Delta\theta\sim
10^{-16}$ and since we know that the values of $\lambda_{30000}$ can
be as small as $10^{-70}$ inside the deepest fjords (and see below), we see that 
\begin{equation}
\frac{\Delta s}{\sqrt \lambda_0}\sim \frac{10^{-16}}{10^{-35}}=10^{19} \ .
\label{resolution}
\end{equation}
Therefore the resolution in the physcial space which is necessary to achieve
a meaningful probe of the deep fjord is highly inappropriate.

The bottom line is that to compute the values of $\lambda_n(s)$ effectively 
we must use the
full power of our iterated conformal dynamics, carrying the history
with us, to iterate forward and backward at
will to resolve accurately the $\theta$ and $\lambda_n$ values 
associated with any given particle on the fully grown cluster.

\begin{figure}
\epsfxsize=6truecm
\epsfbox{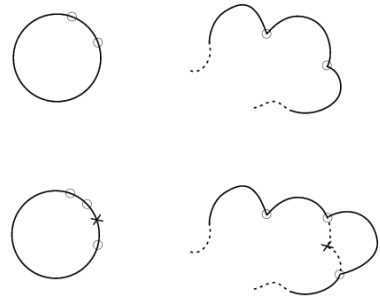}
\caption{A typical growth process in which an existing branch-cut
is ``burried" under the new bump. Such events reduce the number
of branch cuts below $2n$ with $n$ being the number of particles.}
\label{largeX}
\end{figure}

To do this we recognize that every time we grow a semi-circular bump
we generate two new branch-cuts in the map  $\Phi^{(n)}$. We find the
position
on the boundary between every two branch-cuts, and there compute
the value of $\lambda_n$. The first step in our algorithm is to generate
the location
of these points intermediate to the branch-cuts \cite{01BDP}. Each branch-cut has a
preimage on the
unit circle  which will be indexed with 3
indices, $w^{k(\ell)}_{j,\ell}\equiv \exp[i\theta^{k(\ell)}_{j,\ell}]$.
The index $j$ represents the generation when
the branch-cut was created (i.e. when the $j$th particle was grown).
The index $\ell$ stands for the generation at which the analysis is
being done (i.e. when the cluster has $\ell$ particles). The index $k$
represents the position of the
branch-cut along the arc-length, and it is a function of the
generation $\ell$.  Note that
since bumps may overlap during growth, branch-cuts are then
covered, cf. Fig. \ref{largeX}. Therefore the maximal $k$, $k_{max} \le 2\ell$.
After each iteration the preimage of each branch-cut moves on
the unit circle, but its physical position remains. This leads
to the equation
that relates the indices of a still exposed branch-cut that was
created at generation $j$ to a later generation $n$:
\begin{eqnarray}
\Phi^{(n)}(w^{k(n)}_{j,n})&\equiv&
\Phi^{(n)}\left(\phi^{-1}_{\lambda_n,\theta_n}
\circ\dots\circ\phi^{-1}_{\lambda_{j+1},\theta_{j+1}}
(w^{\tilde k(j)}_{j,j})\right)\nonumber\\&=& \Phi^{(j)}(w^{\tilde
k(j)}_{j,j}) \ . \label{trick}
\end{eqnarray}
Note that the sorting indices $\tilde k(j)$ are not simply
related to $k(n)$, and need to be tracked as follows.
Suppose that the list $w^{k(n-1)}_{j,n-1}$ is available. In the $n$th
generation we
choose randomly a new $\theta_n$, and find two new branch-cuts
which on the unit circle are at angles $\theta_n^{\pm}$. If one (or very
rarely more)
branch-cut of the updated list
$\phi^{-1}_{\lambda_n,\theta_n}(w^{k(n-1)}_{j,n-1})$ is covered,
it is eliminated from the
list, and together with the sorted new pair we make the list
$w^{k(n)}_{j,n}$. 

Next let us find the mid-positions at which we want to compute
the value of $\lambda_n$.
Having a cluster of $n$ particles we now consider
all neighboring pairs of preimages $w^{k(n)}_{j,n}$ and
$w^{k(n)+1}_{J,n}$,
that very well may have been created at two different generations $j$
and $J$.
The larger of these indices ($J$ without loss of generality)
determines the generation of the intermediate
position at which we want to compute the field. We want to find the
preimage
$u^{k(n)}_{J,J}$ of this mid-point on the unit circle , to compute
$\lambda_{k(n)}$
there accurately. Using definition (\ref{trick}) we find the preimage
\begin{equation}
\arg(u^{k(n)}_{J,J}) = [\arg(w^{\tilde k(J)}_{j,J})+\arg(w^{\tilde
k(J)+1}_{J,J})]/2 \ .
\end{equation}
In Fig.1 panel b we show, for the same cluster of 50 000, the map
$\Phi^{(J)}(u^{k(n)}_{J,J})$
with $k(n)$ running between 1 and $k_{\rm max}$, with
$J$ being the corresponding generation of creation of the mid point. We
see that
now all the particles are probed, and
every single value of $\lambda_{k(n)}$ can be computed. 

To compute these $\lambda_{k(n)}$
accurately, we define (in analogy to Eq.(\ref{trick})) for every $J<m\le n$
\begin{equation}
u^{k(n)}_{J,m}\equiv \phi^{-1}_{\lambda_m,\theta_m}\circ \dots
\circ
\phi^{-1}_{\lambda_{j+1},\theta_{j+1}}(u^{k(n)}_{J,J}) \ .
\end{equation}
Finally $\lambda_{k(n)}$ is computed from the definition
(\ref{lambdan})
with
\begin{eqnarray}
{\Phi^{(n)}}'(u^{k(n)}_{J,n})
&=&\phi'_{\lambda_n,\theta_n}(u^{k(n)}_{J,n})
\cdots\phi'_{\lambda_{J+1},\theta_{J+1}}(u^{k(n)}_{J,J+1})\nonumber\\&\times&
{\Phi^{(J)}}'(u^{k(n)}_{J,J}) \ . \label{prod}
\end{eqnarray}

We wish to emphasize the relevance of this equation: the problem with the coarse
resolution that was exposed by Eq. (\ref{resolution}) occurs only
inside the deepest fjords. We note however that the particles inside the deep fjords were
deposited when the clusters were still very  small. For small clusters the
resolution of the fjords does not pose a difficult problem. Therefore, when we
evaluate the derivative $\Phi^{(n)}$ inside the deepest fjord at a
point $u_{J,n}^{k(n)}$, we make use of the fact that $J\ll n$ and
write the derivative in the form
\begin{equation}
\label{formula}
\Phi^{(n)'}(u_{J,n}^{k(n)})=\Phi^{(J)'}(u_{J,J}^{k(n)})\times \mbox{``correcting terms''}
\end{equation}
On the left hand side of Eq. (\ref{prod}) we see that within our limited numerical
resolution $u_{J,n}^{k(n)}$, $u_{J,n}^{k(n)+1}$ and the
correponding values of $\lambda_n$ are almost identical
whereas for the righthand side this is not the case. By keeping
track of the branch cuts we improve the precision inside the fjords
dramatically. In other words, the large screening inside the
fjords is simultaneously the problem and the solution. 
The problem is that we
cannot use the standard approach in evaluating $\Phi^{(n)'}$. The
solution is that for a point $u_{J,n}^{k(n)}$ inside the deepest
fjords we always have that $J\ll n$ and therefore the evaluation
(\ref{prod}) helps to improve the resolution. 

In summary, the calculation is optimally accurate since we avoid as much
as possible the effects of the rapid shrinking of low probability 
regions on the unit circle.
Each derivative in (\ref{prod}) is computed using information from
a generation in which points on the unit circle are optimally resolved.

The integral (\ref{mean}) is then estimated as the finite sum
$\sqrt{\lambda_0}\sum_{k(n)}  \lambda_{k(n)}^q$.
We should stress that for clusters of the order of 30 000 particles we
already
compute, using this algorithm, $\lambda_{k(n)}$ values of the order of
$10^{-70}$.
To find the equivalent small probabilities using random walks would
require
about $10^{35}$ attempts to see them just once. This is of course
impossible,
explaining the lasting confusion about the issue of the phase transition
in
this problem. This also means that all the $f(\alpha)$ curves that were
computed before \cite{averages,87HSM} did not converge. Note that in
our calculation the small values of $\lambda_{k(n)}$ are obtained from
multiplications
rather than additions, and therefore can be trusted.
\section{Multifracal analysis of the harmoninc measure}
\begin{figure}
\epsfxsize=7truecm
\epsfbox{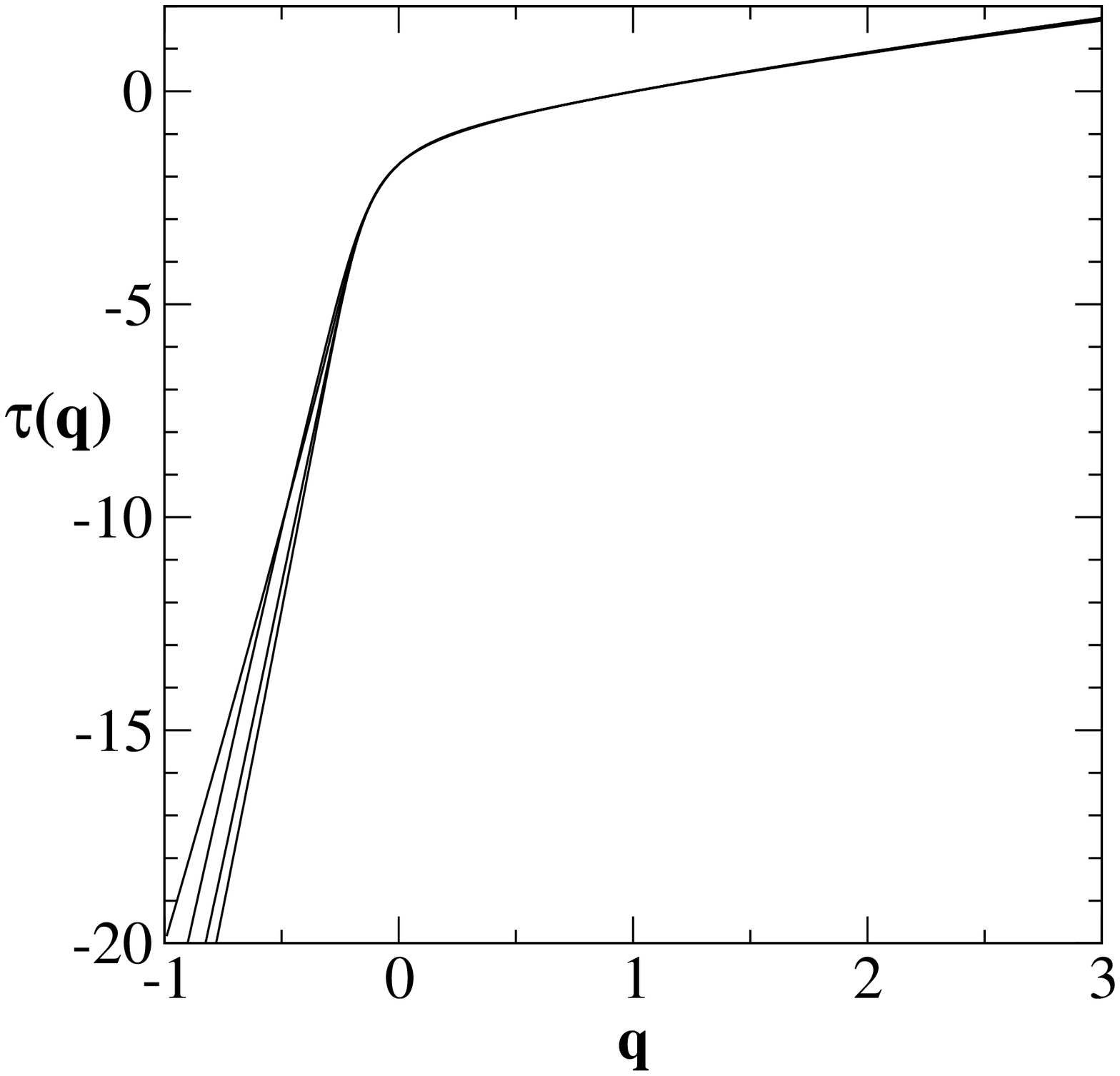}
\epsfxsize=7truecm
\epsfbox{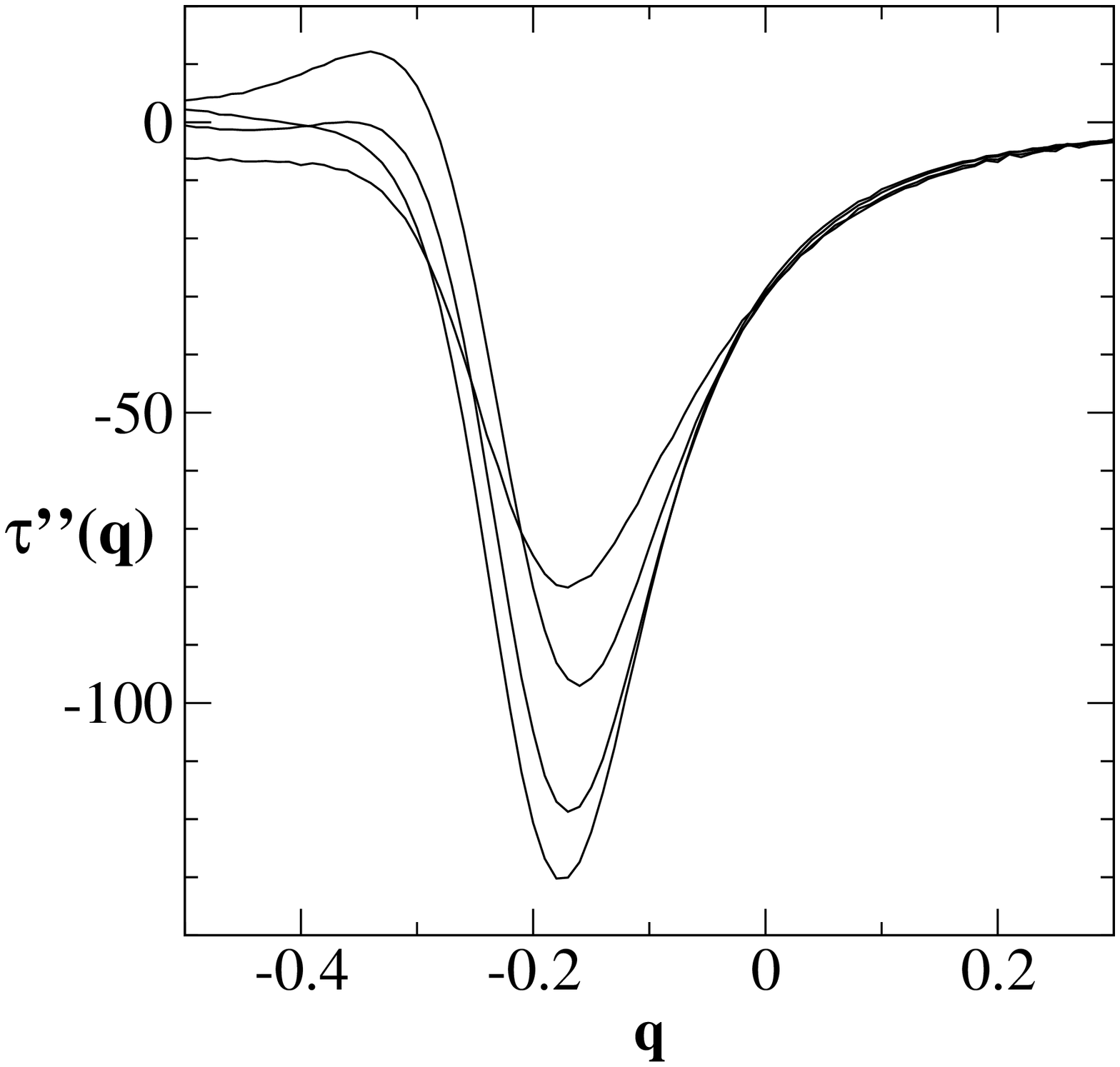}
\caption{Panel a: the calculated function $\tau(q)$ for clusters of $n$
particles,
with $n=$10 000,15 000, 25 000 and 30 000. Panel b: the second
derivative of $\tau(q)$
with respect to $q$.}
\label{tau}
\end{figure}

\begin{figure}
\epsfxsize=7truecm
\epsfbox{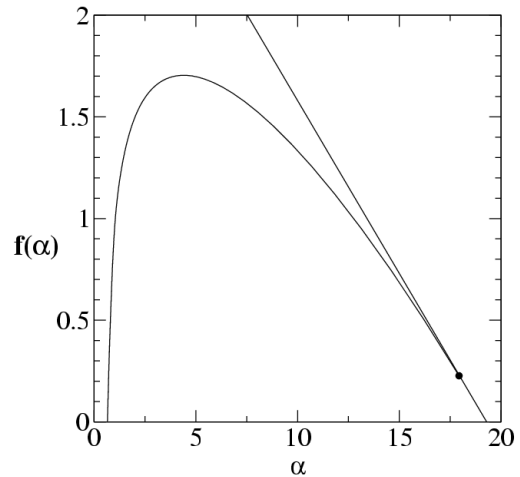}
\caption{The calculated function $f(\alpha)$ using $\tau(q)$ calculated
from a cluster with  $n=30~000$ particles. This $f(\alpha)$ is
almost indistinguishable from the one computed with
$n=25 000$ particles. We propose that this
function is well converged. The black dot denotes where the
curve ends, being tangent to the line with slope -0.18.}
\label{alfalfa}
\end{figure}

Having the accurate values $\lambda_{k(n)}$ we can now compute the moments
(\ref{lamDq}).
Since the scaling form on the RHS includes unknown coefficients, we
compute
the values of $\tau(q)$ by dividing $\langle\lambda_n^q\rangle$
by $\langle\lambda_{\bar n}^q\rangle$, estimating
\begin{equation}
\tau(q) \approx -D \frac{\log{\langle\lambda_n^q\rangle}
-\log{\langle\lambda_{\bar n}^q\rangle}}{\log{n}-\log{\bar n}}
\end{equation}
Results for $\tau(q)$ for increasing values of $n$ and $\bar n$ are
shown in
Fig. \ref{tau}, panel a. 
It is seen that the value of $\tau(q)$ appears to grow
without bound for $q$ negative. The existence of a phase transition is
however best indicated by measuring the derivatives of $\tau(q)$ with
respect
to $q$. In Fig. \ref{tau} panel b we show the second derivative, indicating a
phase
transition at a value of $q$ that recedes {\em away} from $q=0$ when $n$
increases. Due to the high accuracy of our measurement of $\lambda$ we
can estimate already with clusters as small as 20-30 000 the $q$ value of
the phase transition as $q^*=-0.18\pm 0.04$. It is quite possible
that larger clusters would have indicated slighly more negative
values of $q^*$, (and see below the results of different methods of
estimates), but we believe that this value is close to convergence.
The fact that this is so can be seen from the $f(\alpha)$
curve which is plotted in Fig. \ref{alfalfa}.
A test of convergence is that the
slope of this function where it becomes essentially linear must agree
with the $q$ value of the phase transition. The straight line shown
in Fig.3 has the slope of -0.18, and it indeed approximates very
accurately the slope of the $f(\alpha)$ curve where it stops
being analytic. The reader should also note that the peak of the
curve agrees with $D\approx 1.71$, as well as the fact that
$\tau(3)$ is also $D$ as expected in this problem.  The value
of $\alpha_{\rm max}$ is close to 20, which is higher than
anything predicted before. It is nevertheless finite. We believe
that this function is well converged, in contradistinction with
past calculations.
\section{Alternative way to approach the phase transition}

An alternative way to the multifractal analysis is obtained by
first reordering all the computed values of $\lambda_{k(n)}$ in
ascending order. In other words, we write them as a sequence
$\{\lambda_n(i)\}_{i\in I}$ where $I$ is an ordering of the indices
such that $\lambda_n(i)\leq \lambda_n(j)$ if $i<j$. The number of
samples we consider is usually large and therefore the discrete index $i/N$ might be treated
as a continuous index $0\le x\le 1$ and $\lambda_n$ as a non-decreasing function of $x$,
\begin{equation}
\lambda_n\equiv f(x)
\end{equation}

\begin{figure}
\centering
\includegraphics[width=.4\textwidth]{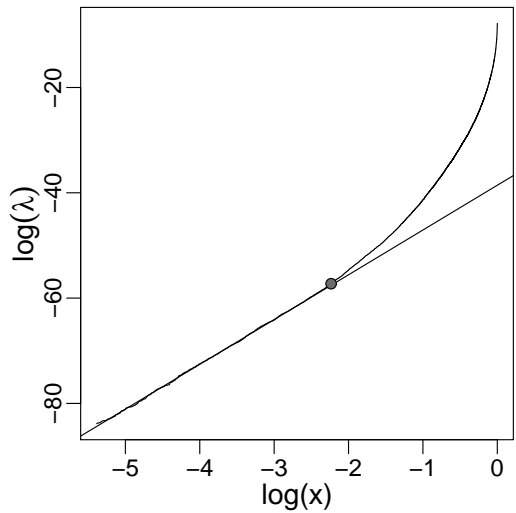}
\caption{Values of $\lambda_n$ sorted in ascending order with respect to the variable 
$x=i/N$. This function is a pure power law for values of $\log x$ smaller than
the position of the circle. The power law is characterized by an exponent
$\beta\approx 8.55$. This is consistent with a phase transition for
$q^*\approx-.23$}\label{lscl}
\end{figure}

We next consider the distribution of $p(\lambda_n)$, which is calculated by
the usual transformation formula
\begin{equation}
p(\lambda_n)\sim\int\delta(\lambda_n-f(x))dx=\frac 1 {|f'(x(\lambda_n))|}
\end{equation}
Using the distribution of $\lambda_n$ we now do the following rewritings,
\begin{eqnarray}
\int_0^{2\pi}\lambda_n^q d\theta&=&\int_0^{L}\lambda_n^q
\frac{d\theta}{ds}ds
\sim\int_0^{L}\lambda_n^{q+\frac 1 2}ds\\
&\sim&\int_0^{\infty}\lambda_n^{q+\frac 1 2}p(\lambda_n)d\lambda_n \ . \label{plam}
\end{eqnarray}
In Fig. \ref{lscl} we will show that our function $f(x)$ obeys a power law
for low values of $x$,
\begin{equation}
f(x)\sim x^\beta \ , \quad {\rm for }~x\ll 1 \ . \label{fscales}
\end{equation}
This in turn implies a power law dependence of $p(\lambda_n)$ on $\lambda_n$
\begin{equation}
p(\lambda_n)\sim\frac 1 {\big(x(\lambda_n)\big)^{\beta-1}}\nonumber
\sim\lambda_n^{\frac {1-\beta}\beta} \ , \quad {\rm for}~ \lambda_n\ll 1 \ .
\label{pscales}
\end{equation}

This power law tail means that the moment integral (\ref{plam}) diverges for values of
$q$ below a critical value $q_c$ given by,
\begin{equation}
q_c+\frac 1 2 +\frac{1-\beta} \beta=-1
\end{equation}
Thus $q_c=-\frac 1 2 -\frac 1 \beta$. From (\ref{lamDq}) we see that the value
of $q^*$ satisfied the relation
\begin{equation}
q^*=2q_c+1=-2/\beta \ . \label{qbeta}
\end{equation}

In Fig. \ref{lscl} we show how the values of $\lambda_n$ depend
on $x$ for small values of $x$. The data is taken from a cluster with $n=20000$.
Denoting the value of $\lambda_n$ that is marked as a full circle
by $\lambda_c$, the figure supports the existence of the power
law (\ref{fscales}) for values of $\lambda_n$ smaller than $\lambda_c$.
Needless to say this alos implies that $p(\lambda_n)$ scales according
to Eq.(\ref{pscales}).
By averaging over 16 clusters of size n=20000 we estimate the
slope in Fig. \ref{lscl} to be $\beta\approx 8.55$ or 
\begin{equation}
q^*=-0.23\pm .05 \ .
\end{equation}
Obviously, this result is in agreement with our direct calculation
in Sect. 3.
\section{Geometrical interpretation of the phase transition}

At this point we would like to interpret the origin of the
phase transition, which in light of the last
section stems from the power law behavior of $p(\lambda_n)$
for small values of $\lambda_n$.
We first identify the region on the cluster that supports the
low values of $\lambda_n$ that belong to the power-law tail of $p(\lambda_n)$.

\begin{figure}
\centering
\includegraphics[width=.4\textwidth]{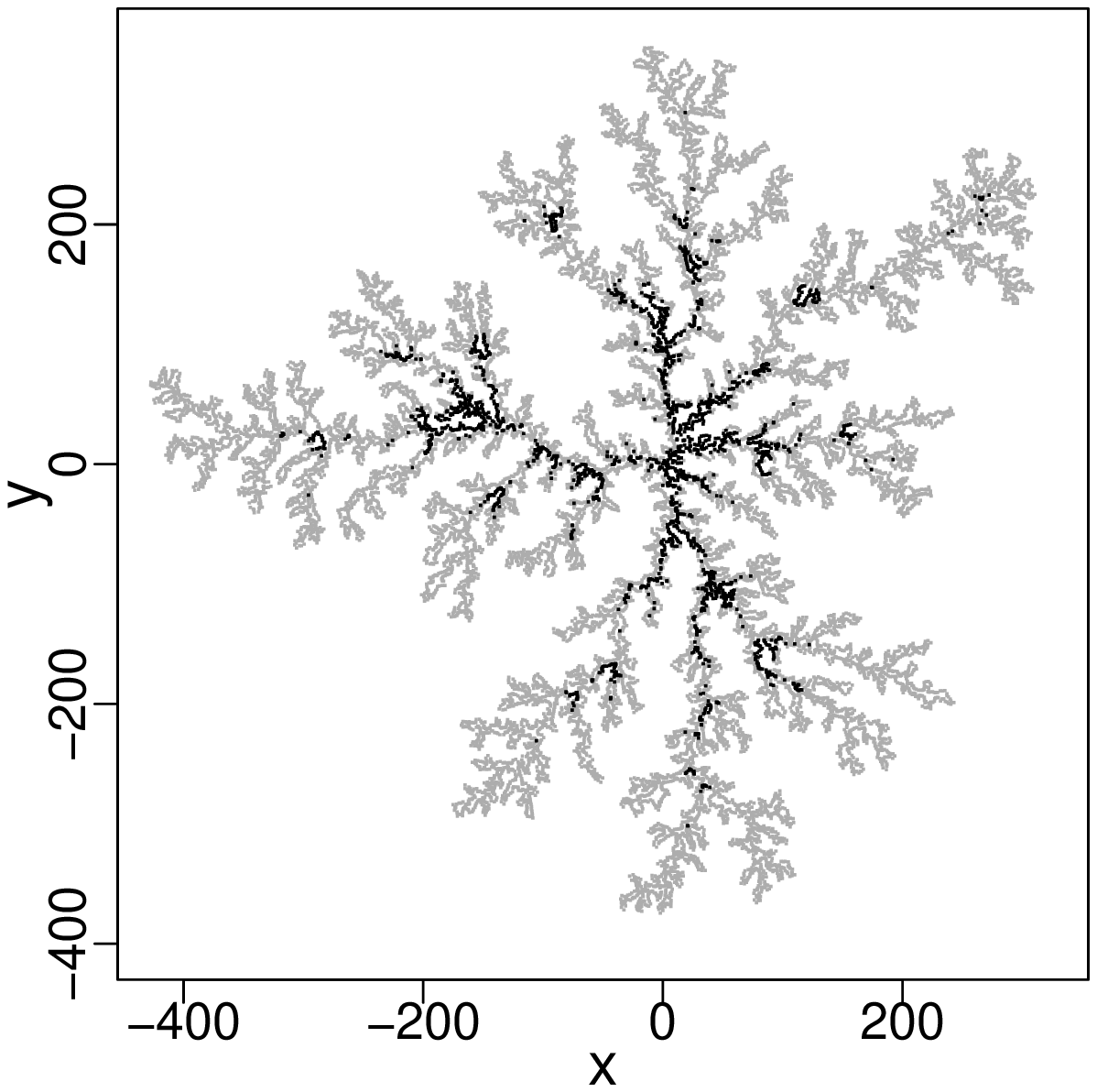}
\includegraphics[width=.4\textwidth]{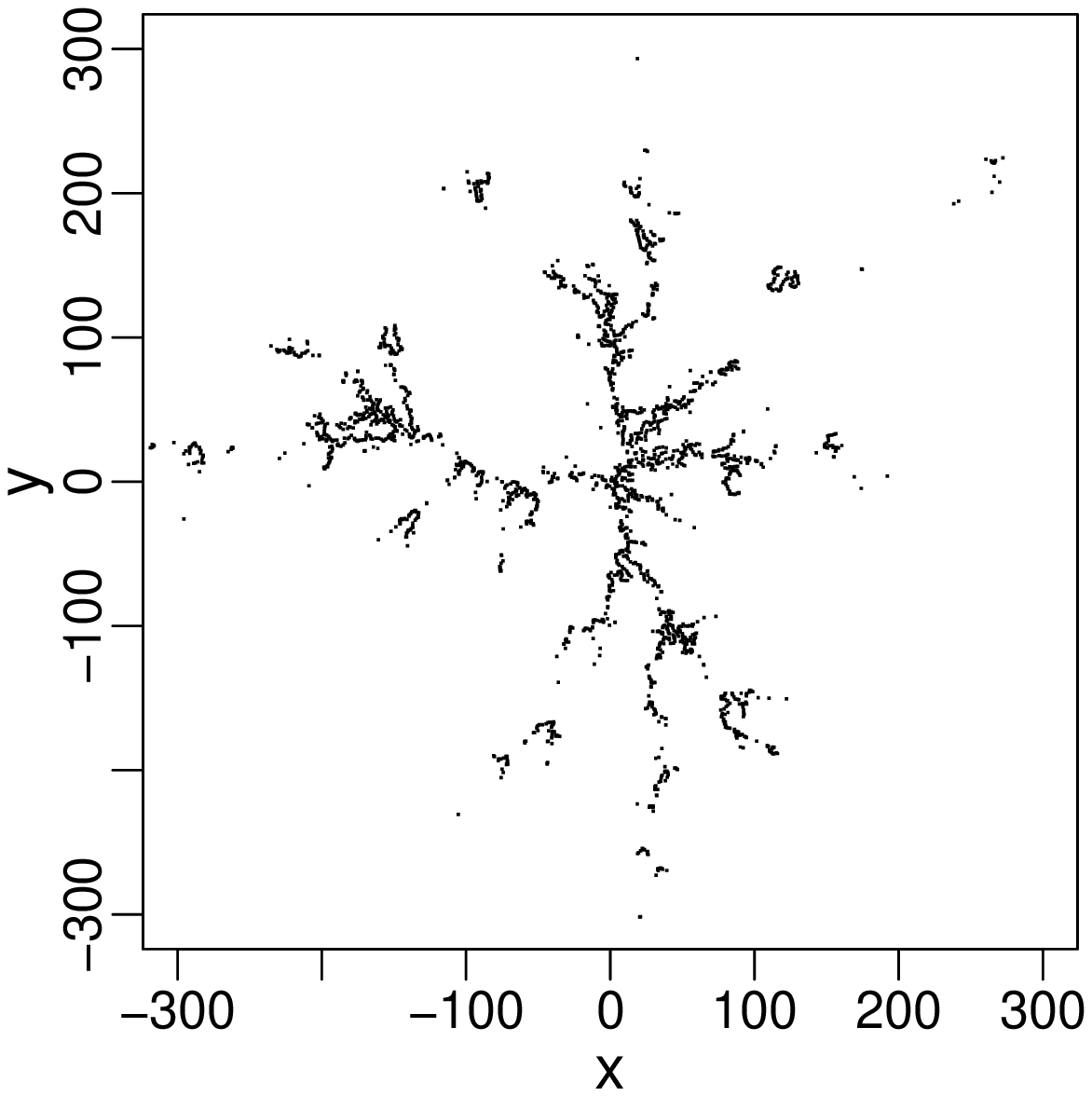}
\caption{The set of all particles which are associated with values
of $\lambda_n$ belonging to the power law region shown in Fig. \ref{lscl}.
In panel a we show the set on the background of the cluster in grey, and in
panel b the set isolated from the rest of the cluster.}
\label{thinset}
\end{figure}

Consider again Fig. \ref{lscl}.
The point denoted above as $\lambda_c$ 
defines the maximum value for which we see a power law in 
$\lambda_n$ vs. $x$. Therefore the set
responsible for the phase transition is the union of bumps with a
value of $\lambda_n$ for which $\lambda_n<\lambda_c$. This set is referred
to below as the ``critical set", and is shown
in figure \ref{thinset}, both on the background of the rest of the cluster,
and as an isolated set.
This figure suggests a geometric interpretation: the fjords in the figure 
all seem to have a
characteristic angle. We will try first to confirm this impression
using careful numerics.

\begin{figure}
\centering
\includegraphics[width=.4\textwidth]{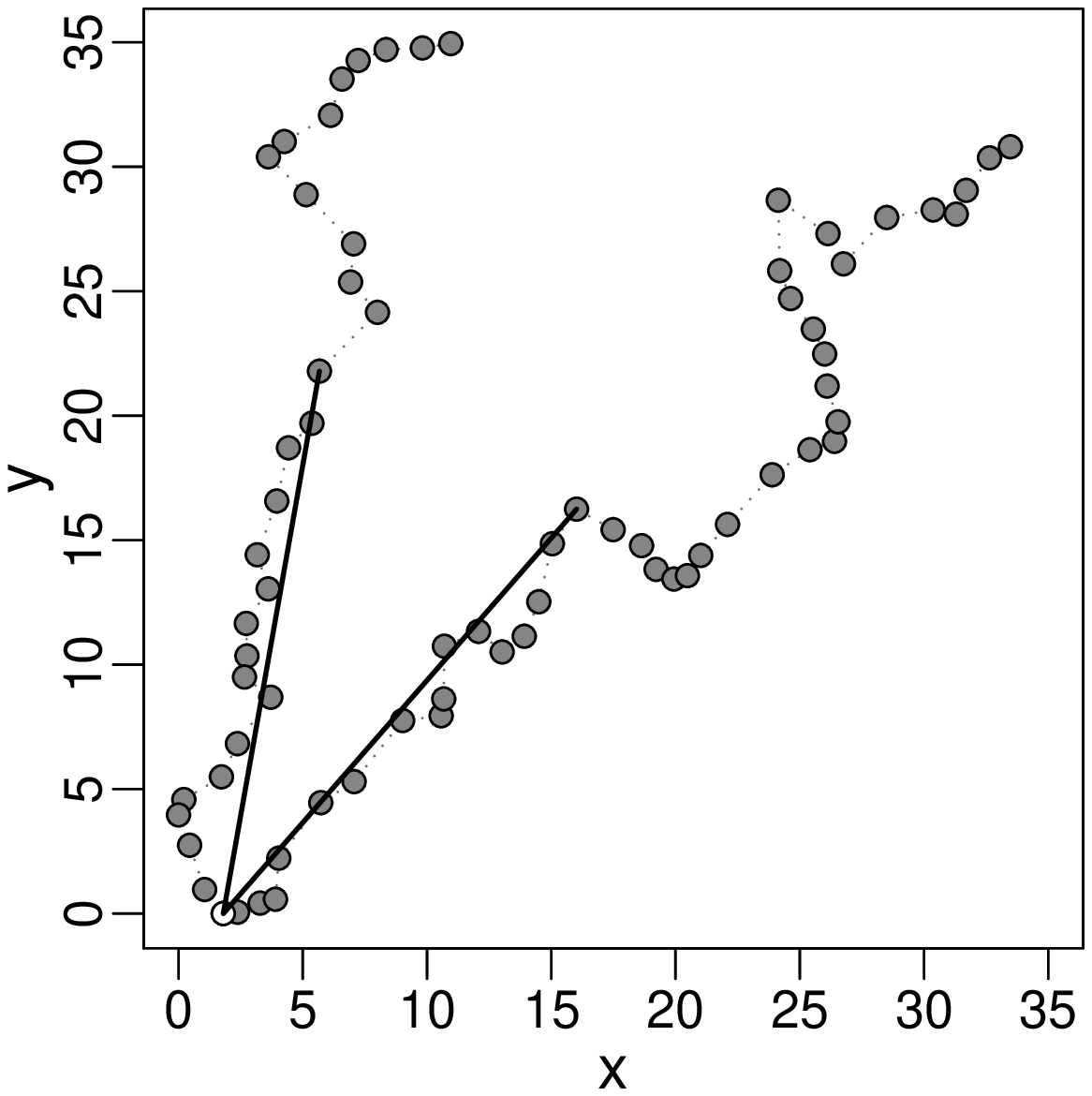}
\includegraphics[width=.4\textwidth]{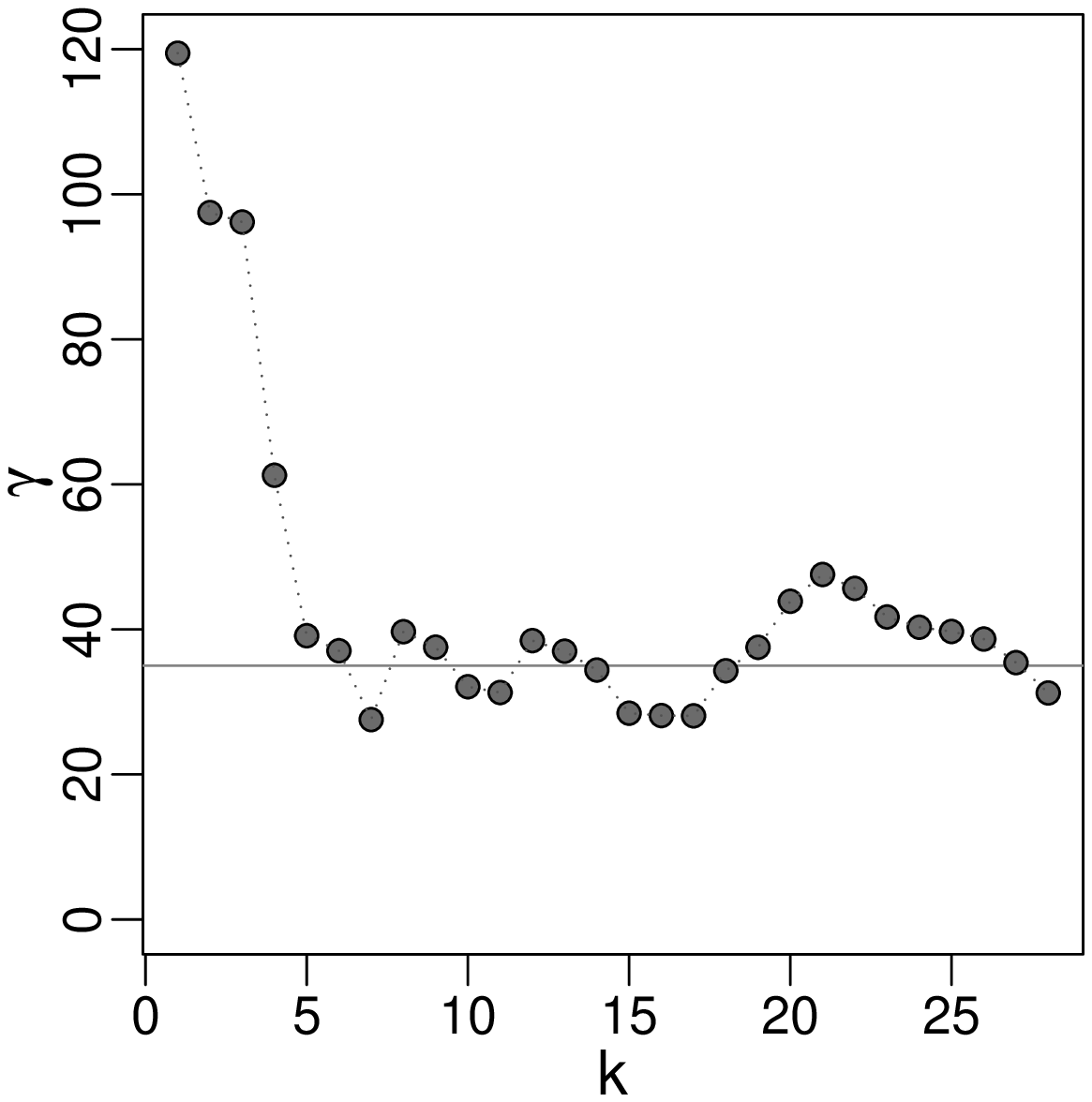}
\caption{Panel a: A typical deep fjord resolved on the sacle of
the particles. From the deepest particle the angle is computed as
explained in the text. Panel b: the change of the measured angle
$\gamma$ as a function of the number of steps $k$ away from
the deepest particle. The angle settles on a value that depends
only weakly on $k$.}\label{angles}
\end{figure}

\begin{figure}
\centering
\includegraphics[width=.4\textwidth]{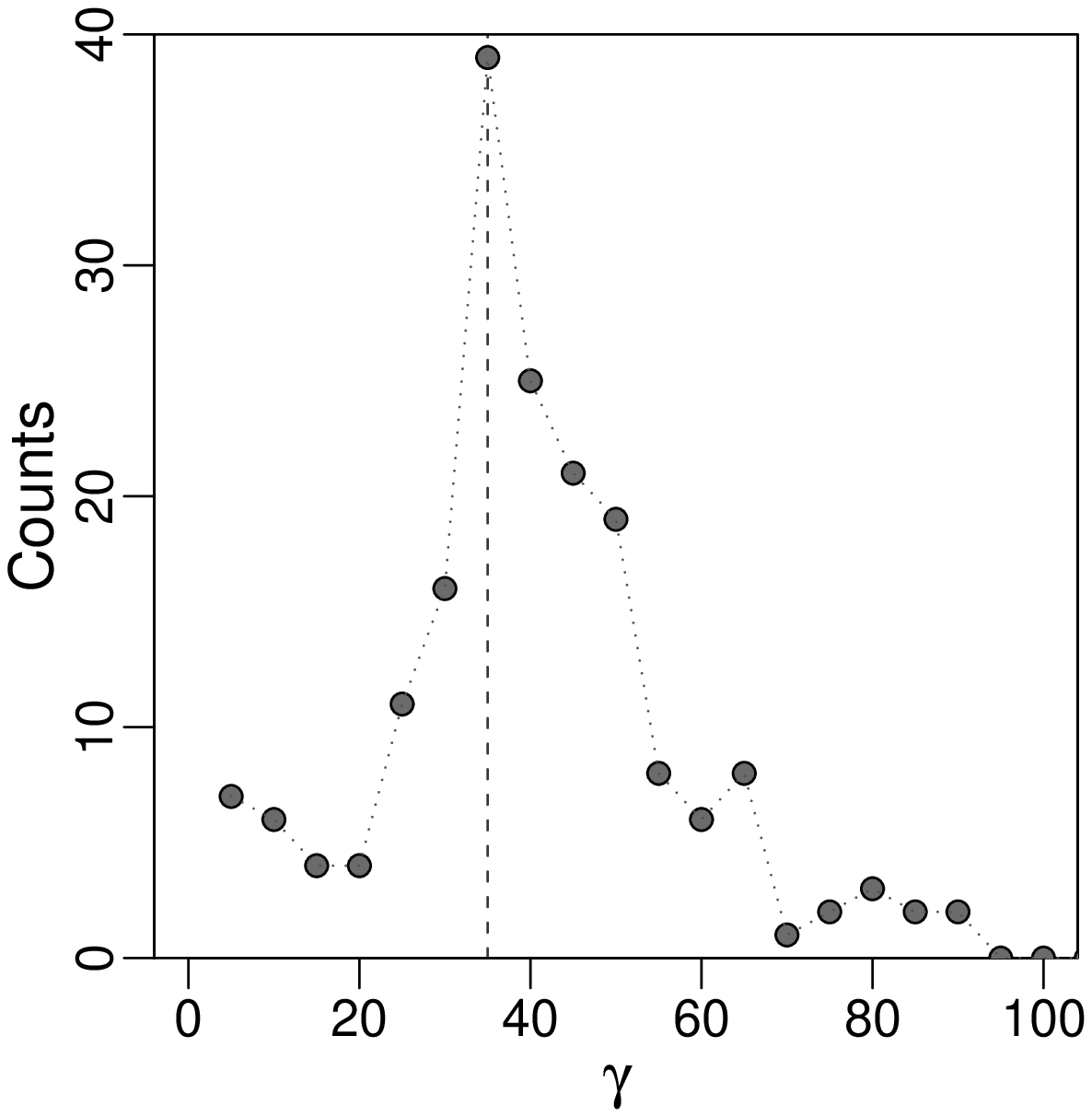}
\caption{The distribution of angles $\gamma$ as determined by the 
procedure exemplified in Fig.\ref{angles} over the set shown in Fig. \ref{thinset}.}
\label{peak}
\end{figure}

Clearly, the set has
several fjords; we  consider them individually. Fig. \ref{angles} shows an example of
such a fjord.  For each fjord we find the point with the minimum probability
and use this for defining the bottom (or deepest point). Second, from the inside, we move to
the two adjacent points which together with the deepest point define an angle. This
angle is recorded, and we move to the next pair of points, and so on
until the value of $\lambda_n$ exceeds $\lambda_c$.  Fig.\ref{angles} 
panel b shows how the angle varies with the number of steps $k$.
For most of the fjords considered the angle is quite large for a small number of steps
(up to 3-4 steps). 
As more steps are taken the angle settles on a characteristic value around
which it fluctuates. For a larger number of steps we reach the
outer parts of the fjord and the angle does no longer reflect
the geometry inside the fjord. The dependence of the angle on $k$ as shown in
Fig. \ref{angles} is typical for all the fjords of the set causing the
phase transition and therefore we see a peak in the distribution
of all the measured angles.  This peak identifies an angle which is
characteristic to the fjords. Fig. \ref{peak} is the distribution
of such typical angles over one cluster. 
We determine the characteristic angle, say $\gamma_c$ by locating the maximum of
the distribution. Finally,
we calculated the average of the charateristic angle $\gamma_c$ over
15 clusters of size $n=20000$. On the basis of that we determine the angle
to be
\begin{equation}
\gamma_c=35^\circ\pm 6^\circ \ . \label{gammac}
\end{equation}

Finally, we can offer a geometrical model to interpret the phase transition.
The results presented in this section indicate that to a reasonable 
approximation the least accessible fjords can be modeled as a wedge of
included angle $\gamma_c$. In Appendix \ref{wedge} we compute the
power law expected for $p(\lambda_n)$ for a wedge. The final result is
\begin{equation}
p(\lambda_n)  \sim  \lambda_n^{-\frac{\alpha-2}{2(\alpha-1)}} \ , \label{poflam} 
\end{equation}
where $\gamma_c=\pi/\alpha$. Using our numerical result for $\gamma_c$ and
Eq.(\ref{qbeta}) we predict finally
$q^*=-.24\pm.05$. Obviously, this is in excellent agreement with the
previous findings.

In addition, we should comment on the interpretation of $f(\alpha_{max})$ which
is the value of the $f(\alpha)$ curve at the point of loss of
analyticity. Within the wedge model offered here, this must be the 
fractal dimension of the set of wedges that support the scaling law
(\ref{poflam}). We have attempted to determine this dimension numerically
by counting the number of fjords in the critical set shown in Fig. (\ref{angles})
as a function of the number of paricles $n$ in the cluster. While the result
of such a calculation is consistent with the proposition, the available
statistics is not sufficient to establish it firmly. We thus conclude
with the proposition as a conjecture, i.e. that $f(\alpha_{max})$ can be
interpreted as the dimension of the set of fjords that belong to the
critical set.
\section{Concluding Remarks}
In conclusion, we have used the method of iterated conformal maps
to compute accurately the harmonic measure of DLA clusers of moderate
size. We have explained that we must use the full power of the method
in order to overcome the strong contraction of the regions on the
unit circle that belong to the deep fjords. By iterating back and
forth, using the fact that we own the history of the iteration scheme,
we could resolve probabilities as small as $10^{-35}$. Using this data
we could establish beyond doubt that the generalized dimensions (or,
equivlently, the $f(\alpha)$ function) lose analyticity at a negative value
of $q$, implying the existence of $\alpha_{\rm max}$. In order to understand
the loss of analyticity we offer a geometric picture. We identified the critical
set on the cluster as having harmonic probabilities that belong to the
power law tail of $p(\lambda_n)$. Considering this set we identified
fjords that can be modeled as wedges of characteristic angle. Taking such
wedges as a model for the fjords of the critical set, we found a value
of $q^*$ which is very close to the one computed using other methods.
We thus propose that the point of non-analyticity can be interpreted as
resulting from the power-law dependence of the harmonic measure in
the fjords belonging to the critical set. 

%
%

\acknowledgments
This work has been supported in part by the
Petroleum Research Fund, The  European Commission under the
TMR program and the Naftali and Anna
Backenroth-Bronicki Fund for Research in Chaos and Complexity. A. L.
is supported by a fellowship of the Minerva Foundation, Munich, Germany.
\appendix
\section{Wedge Model for the Fjords of the Critical Set}
\label{wedge}
\subsection{The Conformal Map and the Electric Field}
The conformal function\\
\begin{equation}\label{map}
\chi(w) = \left( i \frac{w+1}{w-1} \right)^{1/\alpha}
\end{equation}
maps the outside of the unit circle to the inside of a wedge with opening angle $\gamma_c =
\pi/\alpha$, where $\alpha \ge 1$ allows $\gamma_c$ be vary between $0$ and $\pi$.\\
To calculate the electric field $E$ we need the inverse of $\chi$:\\
\begin{equation}\label{map_inv}
\chi^{-1}(z) = \frac{z^\alpha+i}{z^\alpha-i}
\end{equation}
From here we see that $\chi^{-1}(0)=-1$ and\\
\begin{equation}
\Phi^{-1}\left( \rho \; e^{i\pi(1\pm1)/2\alpha} \right) \longrightarrow 1
\mbox{\hspace{1cm} as } \rho \rightarrow \infty 
\end{equation} 
Thus the unit circle is unfolded onto the wedge; shifting the point $w=-1$ to the origin
$z=0$; and rotating and stretching the upper half circle onto the real axis and the lower
half circle to the other ray of the wedge $\rho \; e^{i\pi/\alpha}$. 
The electric field follows from its definition:\\
\begin{equation}
E(z) = \left| \frac{d}{dz} \log \left( \chi^{-1}(z) \right)\right| = \frac{2\alpha}{|z|} \;
\frac{1}{\left| z^\alpha + z^{-\alpha}\right|}
\end{equation}
On the real axis close to the center $z=0$, i.e. for $z = \rho \ll 1$ it becomes\\
\begin{equation}\label{E_low}
E(\rho) = \frac{2\alpha}{\rho} \; \frac{1}{\rho^\alpha + \rho^{-\alpha}} \approx 2 \;
\alpha \; \rho^{\alpha-1}
\mbox{\hspace{1cm}for } \rho \ll 1
\end{equation}
while for large $\rho$ is goes like\\
\begin{equation}\label{E_high}
E(\rho) = \frac{2\alpha}{\rho} \; \frac{1}{\rho^\alpha + \rho^{-\alpha}} \approx 2 \;
\alpha \;\rho^{-(\alpha+1)}
\mbox{\hspace{1cm}for } \rho \gg 1 
\end{equation}
Exactly the same relations hold for the upper ray.
\subsection{The Probability Measure for $\lambda_n$}
The linear size of the particles in mathematical space $\sqrt{\lambda_n}$ is proportional
to the electric field.\\
\begin{equation}
\sqrt{\lambda_n}(\theta) = \sqrt{\lambda_0} \; E\left( \Phi \left( e^{i\theta} \right)
\right) 
\end{equation}
Thus the probability measure of the $\lambda_n$ is directly related to the probability
measure of the electric field. Since $E$ is the same for the two rays of the wedge, it is
sufficient to consider it only on the real axis. Starting from the uniform distribution of
the $\theta$-values in mathematical space, it follows:\\
\begin{eqnarray}
1 &=& \frac{dP\left(\theta\right)}{d\theta} = \frac{dP(E)}{dE} \; \left|
\frac{dE}{d\rho} \right| \;  \left|\frac{d\rho}{d\theta}\right| \nonumber\\
&=& \frac{dP}{dE} \;
\left|\frac{dE}{d\rho} \right| \; \left| -i \frac{d}{d\rho}
\log\left(\Phi^{-1}(\rho)\right) \right|^{-1} \nonumber\\
&=& \frac{dP}{dE} \; \left|\frac{dE}{d\rho} \right| \; E^{-1} 
\end{eqnarray}
or\\
\begin{equation}\label{E_prob}
dP(E) \sim \left| \frac{dE}{d\rho} (\rho) \right|^{-1} \; E \; dE
\end{equation}
The derivative of the electric field follows from (\ref{E_low}) or (\ref{E_high})\\
\begin{equation}
\left| \frac{dE}{d\rho} (\rho) \right| = \frac{2\alpha}{\rho} \frac{\left| (\alpha+1)
\rho^\alpha - (\alpha-1) \rho^{-\alpha}
\right|}{\left( \rho^\alpha + \rho^{-\alpha}\right)^2} 
\end{equation}
Yielding\\
\begin{equation}
dP(E) \sim \frac{\rho\left(\rho^\alpha + \rho^{-\alpha}\right)}{\left| (\alpha+1)
\rho^\alpha - (\alpha-1)\rho^{-\alpha} \right|}\; dE 
\end{equation}
For small $\rho$ corresponding to a small field and thus small $\sqrt{\lambda_n}$ we
get:
\begin{equation}
dP(E) \sim \rho  dE \sim E^{\frac{1}{\alpha-1}} 
\end{equation}
or
\begin{equation}
dP\left( \sqrt{\lambda_n} \right) \sim \sqrt{\lambda_n}^{-\frac{1}{\alpha-1}}
d\sqrt{\lambda_n}
\end{equation}
For the probability density of $\lambda_n$ this means\\
\begin{eqnarray}\label{lamb_prob}
dP(\lambda_n) & = & \frac{dP\left( \sqrt{\lambda_n}
\right)}{d\sqrt{\lambda_n}} \; \frac{d\lambda_n}{2 \; \sqrt{\lambda_n}}\\
& \sim & \lambda_n^{-\frac{\alpha-2}{2(\alpha-1)}} d\lambda_n
\end{eqnarray}

%
%

\end{document}